\documentstyle[prl,aps]{revtex}
\newcommand{\BEQ}{\begin{equation}}
\newcommand{\EEQ}{\end{equation}}
\newcommand{\BEA}{\begin{eqnarray}}
\newcommand{\EEA}{\end{eqnarray}}

\newcommand{\x}{{\bar x}}
\renewcommand{\r}{{\bar r}}
\newcommand{\q}{\bar q}
\renewcommand{\d}{{\rm d}}
\newcommand{\E}{{\tilde E}}
\begin{document}
 \draft  
\title{ Variational approach to interfaces in random media:  
negative variances and replica symmetry breaking }
\author{D.B. Saakian$^{(1)}$ and Th.~M.~Nieuwenhuizen$^{(2)}$}
\address{
$^{(1)}$ {\em Yerevan Physics Institute}\\
Alikhanian Brothers Street 2, Yerevan 375036, Armenia\\
Saakian @ jerewan1.YERPHI.AM \\
$^{(2)}$ Van der Waals-Zeeman Institute, Universiteit van Amsterdam\\ 
	 Valckenierstraat 65, \\1018 XE Amsterdam, The Netherlands\\
	nieuwenh@phys.uva.nl}
 \date{28-3-97, revised 23-6-97}
\maketitle

\begin{abstract}
A Gaussian variational approximation  is often used to study
interfaces in random media. By considering the 1+1 dimensional
directed polymer in a random medium, it is shown here that the 
variational Ansatz typically leads to a negative variance of 
the free energy. The situation improves by taking into account
more and more steps of replica symmetry breaking. For infinite
order breaking the variance is zero (i.e. subextensive).

This situation is reminiscent of the negative entropies in
mean field spin glass models, which were also eliminated by
considering infinite order replica symmetry breaking.
\end{abstract}
\pacs{7510.Nr,7540.Cx, 7550.Lk,8710}

\subsection*{Running title: Negative variances and replica symmetry breaking}

The subject  of interfaces in random media has many application, such as
directed polymers in a random medium, domain walls is dirty magnets,
and  interfaces in random sponges. 
Various aspects of this field and methods of approach have been reviewed, 
see \cite{NV},\cite{FLN}, \cite{HHZ}.

As a tool to get a grip on these complicated
objects, a Gaussian variational Ansatz 
has been proposed by M\'ezard and Parisi~\cite{MP}
to calculate the free energy.
It was found that replica symmetry is broken spontaneously. This
variational replica-approach reproduces the Flory values for 
critical exponents, that can be derived by Flory-Imry-Ma
type of estimates. 

In the approach an infinity ($k=\infty$) of replica symmetry breakings occur.
Carlucci, de Dominicis and Temesvari have shown that the 
instability of the replica approach at finite $k$ disappear in the limit 
$k\to\infty$.~\cite{CdDT}

The problem of  a directed polymer on a random substrate in $d=1+1$
has received quite some attention. It was pointed out by Kardar~\cite{Kardar85}
that the replicated free energy  can be related to the ground state energy of
$n$ interacting bosons in one dimension. This problem can be solved exactly 
by means of the Bethe Ansatz. 
The depinning of such a polymer or interface from a wall 
is also solvable exactly, when the effect of the wall potential 
is described  by a derivative 
condition of the wavefunction at the wall. Various interesting results 
have been
derived from the exact groundstate wavefunction near the depinning transition.

Here we reconsider the $d=1+1$ problem of a directed polymer of
length $L$ in a random potential. 
The purpose is not to derive the best variational approximation
to the exact solution. Having in mind more complicated models, 
we wish to test a particular
 aspect of the variational approach of M\'ezard and Parisi,
 that we fully work out for this situation. Due to the one-dimensional
character of the problem, we can get the results by studying
a quantum Hamiltonian.

In particular we point out that order
$n^2L$ terms of the $n$-fold replicated free energy $F_n$ 
tend to have the wrong sign, so that
they tend to predict a negative variance of the physical free energy. 
This unphysical prediction occurs in case of no replica 
symmetry breaking and becomes less severe for a finite number of breakings;
it seems related to the instability of some fluctuation modes.
The whole $n^2L$ term disappears  
if infinite order replica symmetry breaking is taken  into account. 
The negative variance of the free energy thus plays a 
role similar to the negative entropy of the SK-model: both vanish only when
infinite  replica symmetry breaking is taken into account.

The implication of vanishing of the $n^2L$ term is that the fluctuations
in the free energy are smaller than $L^{1/2}$.
We shall then show that the 
replicated free energy has leading correction $\sim n^5L$, which means
that the free energy fluctuations are of predicted to be of
 order $L^{1/5}$. This is
in agreement  with the Flory prediction worked out by M\'ezard
and Parisi. Note that the exact order of
magnitude is known to be $L^{1/3}$.

In section 1 we introduce the model and the variational Ansatz.
The simplest  cases are considered.
In section 2 the problem is solved for infinite order replica 
symmertry breaking. We close with a summary in section 3.

\section{The model and the variational approach}
We consider an interface $z(x)$ with Hamiltonian
\BEQ
\beta{\cal H}=\int_0^L \d x
\{\frac{\gamma}{2}\left(\frac{\d z}{\d x}\right)^2+V(x,z(x))\}
\EEQ
$\beta=1/T$ is the inverse temperature, $\gamma(T)$ is the interface
stiffness, and there is $\delta$-correlated Gaussian disorder
\BEQ
\overline{V(x,z)V(x',z')}=\sigma^2\delta (x-x')\delta(z-z')
\EEQ
where the overline denotes the average over the quenched disorder.
We shall  study the  partition sum
\BEQ
Z=\int Dz\,\,e^{-\beta {\cal H}}=e^{-\beta F}
\EEQ
(For a formulation on the lattice, see ~\cite{FLN}). 
The replicated partition sum 
\BEQ
Z_n=\overline{Z^n}\qquad\longrightarrow\qquad
\exp(-\beta F_n)=\overline{\exp(-n\beta F)}=\exp\sum_{k=1}^\infty
n^k\overline{(-\beta F)^k_{cum}}
\EEQ 
generates the cumulants of $\beta F$ as the expansion coefficients of 
$\beta F_n$ for  small $n$. For having a non-negative variance 
of $F$ the $n^2$ term of $F_n$ should be non-positive.

Averaging over disorder one obtains the effective $n$-particle Hamiltonian
\BEQ 
{\cal H}_n=\int_0^L \d x\{\frac{\gamma}{2}\sum_{\alpha=1}^n
(\frac{\d z_\alpha}{\d x})^2
% \frac{\partial^2}{\partial z_\alpha^2}
-\sigma^2\sum_{\alpha<\beta} \delta(z_\alpha(x)-z_\beta(x))\}
\EEQ

By  analog with the Feynman-Kac path integral formulation of quantum mechanics,
the replicated free energy can be obtained as 
\BEQ \beta F_n=L E_n\EEQ 
where
$E_n$ is the  ground state energy of the quantum Hamiltonian
\BEQ
 H_n=-\frac{1}{2\gamma}\sum_{\alpha=1}^n\frac{\partial ^2}
{\partial z_\alpha^2}
-\sigma^2\sum_{\alpha<\beta}\delta(z_\alpha-z_\beta)
\EEQ
The expansion 
\BEQ
 E_n=nE^{(1)}+n^2E^{(2)}+n^3E^{(3)}+n^4E^{(4)}+n^5E^{(5)}+\cdots
\EEQ
determines the cumulants
\BEQ
\overline {(\beta F)^k_{cum}}=(-1)^{k-1}k!L E^{(k)}
\EEQ

The present problem has an exact ground state wavefunction 
\BEQ \psi_0(z)=\exp-\kappa\sum_{\alpha<\beta}|z_\alpha-z_\beta|
\EEQ
where $\kappa=\sigma^2\gamma$. The groundstate energy is
\BEQ
 E_n=\frac{\gamma \sigma^4}{2}\,\frac{n-n^3}{3}
\EEQ
To our knowledge, the variational approach 
to this problem was first formulated by
Honeycutt and Thirumalai~\cite{HonThir}. They considered a 
Hartree variational interface Hamitonian, that we generalize as
\BEQ 
{\cal H}_n^{\rm var}= \int_0^L\d x\{
\frac{\gamma}{2}\sum_{\alpha=1}^n
\left(\frac{\d z_\alpha}{\d x}\right)^2
-\frac{\gamma}{2}\sum_{\alpha,\beta=1}^n z_\alpha (q^2)_{\alpha\beta} z_\beta\}
\EEQ
The original approach had $q_{\alpha\beta}=\delta_{\alpha\beta} q$.
 The generalized
approach is equivalent with the one of M\'ezard and Parisi, though it 
is simpler.
Indeed, the mapping onto a quantum problem, which can only be done in $d=1+1$,
 prevents the introduction of a  Fourier wavenumber.  

The variational Hamiltonian is equivalent with considering
 the Hartree variational quantum Hamitonian
\BEQ 
H_n^{\rm var}= 
-\frac{1}{2\gamma}\sum_{\alpha=1}^n
\frac{\partial^2}{\partial z_\alpha^2}
-\frac{\gamma}{2}\sum_{\alpha,\beta=1}^n 
z_\alpha (q^2)_{\alpha\beta} z_\beta
\EEQ
which has the groundstate wavefunction
\BEQ
\psi(\{z_\alpha\})=\frac{1}{({\rm det}\pi q)^{1/4}}
e^{-\frac{1}{2}\sum_{\alpha,\beta=1}^n z_\alpha q_{\alpha\beta} z_\beta}
\EEQ

The ground state energy now is approximated as 
\BEA
E_n =E_n^{\rm var}+\langle\psi| (H_n- H_n^{\rm var})|\psi\rangle
\EEA
Rescaling $q\to 2\sigma^4\gamma^2/\pi$ we have
\BEQ
E_n=\frac{\gamma\sigma^4}{2\pi}\tilde E_n\EEQ
with
\BEQ
\tilde E_n=\sum_\alpha q_{\alpha\alpha}
%-\frac{\sigma^2}{4\sqrt{\pi}}
-\sum_{\alpha\neq \beta}\frac{2\sqrt{2}}{\sqrt{
(q^{-1})_{\alpha\alpha}+(q^{-1})_{\beta\beta}
-2(q^{-1})_{\alpha\beta}}}
\EEQ
The exact value would correspond to $\E_n=\pi(n-n^3)/3$.
We shall now consider various choices for the shape of $q_{\alpha\beta}$.

\subsection{Diagonal $q$}

The case $q_{\alpha\beta}=q_d\delta_{\alpha,\beta}$ was studied 
by Honeycutt and Thirumalai. Optimizing $E_n$ yields
 $q_d=(n-1)^2$, so that
\BEQ 
\tilde E_n = -n(n-1)^2
\EEQ
which differs in sign from the Honeycutt-Thirumalai result.
Since our $n^2$ term is positive, we conclude that 
this result leads to a negative variance of $F$, and is not acceptable.

\subsection{Translational invariance and off-diagonal $q$}

\subsubsection{Replica symmetry}
If there is no RSB then we can set $q_{\alpha\beta}
=(q_d-q_0)\delta_{\alpha,\beta}+q_0$. 
\BEQ 
\tilde E_n=nq_d-{2n(n-1)}{\sqrt{q_d-q_0}}
\EEQ
The structure of this expression shows that we only get a well 
posed problem if we relate $q_0$ and $q_d$ in some way.
Above we considered $q_0=0$ but the result was unsatisfactory.
As we have averaged over disorder, there is translational invariance.
One would expect that the physically relevant variational Ansatz reflects
the translational invariance of the replicated problem. Hereto we 
have to impose
\BEQ 
\sum_\beta q_{\alpha\beta}=0
\EEQ
This condition is equivalent to staying in the ``replicon'' 
or ``ergodon'' subspace.

 implying here that $q_d+(n-1)q_0=0$.
 This allows to eliminate $q_d$, after which we get
\BEQ \E_n =-n(n-1)q_0-2n(n-1)\sqrt{-nq_0}\EEQ
The saddle point is $q_0=-n$, yielding
\BEQ \E_n=-n^2(n-1)\EEQ 
This result has $E_0=0$ and variance $-1$, in both respects better
than previous case. But also this result leads to a negative variance,
and is not acceptable.

\subsubsection{One step replica symmetry breaking}

Now we choose
\BEQ 
q_{\alpha\beta}=(q_d-q_1)\delta_{\alpha\beta}+
(q_1-q_0){\cal E}^{(x_1)}_{\alpha\beta}+q_0 {\cal E}^{(n)}_{\alpha\beta}
\EEQ
where ${\cal E}^{(x_1)}$ is the  Parisi 1RSB matrix having $n/x_1$ diagonal 
$x_1\times x_1$ blocks with elements equal to unity. (Thus ${\cal E}^{(1)}$
is the identity matrix, while ${\cal E}^{(n)}$ 
is a matrix with all elements equal to unity.)
Due to translational invariance we may eliminate $q_d$ and get
\BEQ
q_d=-(x_1-1)q_1-(n-x_1)q_0
\EEQ
The inverse reads
\BEQ 
q^{-1}_{\alpha\beta}=\frac{\delta_{\alpha\beta}}{q_d-q_1}
-\frac{q_1-q_0}{(q_d-q_1)(q_d-(1-x_1)q_1-x_1q_0)}{\cal E}^{(x_1)}_{\alpha\beta}
-\frac{q_0}{(q_d-(1-x_1)q_1-x_1q_0)(q_d-(1-x_1)q_1-x_1q_0+nq_0)}
{\cal E}^{(n)}_{\alpha\beta}
\EEQ
The last factor diverges in case of translational invariance. 
Fortunately, it generally cancels from our expression for $\E_n$. We thus obtain
\BEQ
\tilde E_n=n(1-x_1)q_1+n(x_1-n)q_0+2n(1-x_1)\sqrt{-x_1q_1+(x_1-n)q_0}
+2n(x_1-n)\sqrt{\frac{(-x_1q_1+(x_1-n)q_0)(-nq_0)}{-q_1+(1-n)q_0}}
\EEQ
This has the same $\sqrt{-nq_0}$ behavior as above. Setting 
\BEQ
q_0=-nr
\EEQ
we get up to order $n^2$
\BEQ
\E_n=n(1-x_1)q_1-n^2x_1r+2n(1-x_1)\sqrt{-x_1q_1}
-n^2\frac{x_1(1-x_1)}{\sqrt{-x_1q_1}r}+2n^2x_1\sqrt{x_1r}
\EEQ
Optimization yields $q_1=-x_1+{\cal O}(n)$, so
\BEQ
\E_n=nx_1(1-x_1)-n^2r+2n^2x_1\sqrt{x_1r}
\EEQ
Further optimization fixes  $x_1=1/2$, $r=x_1^3=1/8$, which finally yields
\BEQ
\E_n=\frac{1}{4}n+\frac{1}{8}n^2
\EEQ 
The $n^2$ term is smaller that without replica symmetry breaking but still of
wrong sign. We must allow for more breakings.

\section{Infinite order replica symmetry breaking}

Now let us extend  the above approach and consider k-step RSB.
Let us consider the following representation for a matrix q:
\BEQ q_{\alpha\beta}=(q_d-q_k)\delta_{\alpha\beta}
+\sum_{i=0}^k(q_i-q_{i-1}){\cal E}^{(x_i)}_{\alpha\beta}\EEQ
where $q_{-1}\equiv 0$, $x_j<x_{j-1}$ and $x_{k+1}=1$, ${x_0}=n$.
It thus holds ${\cal E}^{(x_i)}{\cal E}^{(x_k)}=x_j{\cal E}^{(x_k)}$
when $x_j<x_k$. The eigenvalues of $q_{\alpha\beta}$ are
\BEA
 \lambda_k&=&q_d-q_k\nonumber\\
\lambda_i&=&q_d-x_{i+1}q_i-\sum_{j=i+1}^k(x_{j+1}-x_j)q_j\qquad i=0,\cdots,k-1
\nonumber\\
\lambda_{-1}&=&nq_0+\lambda_0 
\EEA
Translational invariance implies $\lambda_{-1}=0$.
 Expressing $q$ in the $\lambda$'s 
we get
\BEQ
q_{\alpha\beta}=\lambda_k\delta_{\alpha\beta}
+\sum_{i=0}^k\frac{\lambda_{i-1}-\lambda_i}{x_i}
{\cal E}^{(x_i)}_{\alpha\beta}
\EEQ
Its inverse has the same structure, so it can be written
\BEA
q_{\alpha\beta}^{-1}\equiv b_{\alpha\beta}&=&(b_d-b_k)\delta_{\alpha\beta}
+\sum_{i=0}^k(b_i-b_{i-1}){\cal E}^{(x_i)}_{\alpha\beta}
\nonumber\\
&=&\frac{1}{\lambda_k}+
\sum_{i=0}^k\frac{1/\lambda_{i-1}-1/\lambda_i}{x_i}{\cal E}^{(x_i)}
\EEA
Let us define
\BEQ
r_{\alpha\beta}=(q^{-1})_{\alpha\alpha} - (q^{-1})_{\alpha\beta}
\EEQ
It has elements $r_i$ 
\BEQ\label{ri=}
r_i=\frac{1}{\lambda_k}+
\sum_{j=i+1}^k\frac{1}{x_j}(\frac{1}{\lambda_{j-1}}-\frac{1}{\lambda_j})
\EEQ
The replicated energy becomes
\BEA \label{En1}
\frac{\E}{n}&=&(x_1-n)q_0+\sum_{i=1}^{k}(x_{j+1}-x_i)q_i
+2\frac{x_0-n}{\sqrt{r_0}}
+2\sum_{i=1}^{k} \frac{x_{i+1}-x_i}{\sqrt{r_i}}
\EEA
where $x_{k+1}\equiv 1$.
This can be rewritten as
\BEA \label{En2}
\E_n=n\E^{(1)}+n^2\Delta_n
\EEA
with 
\BEA \label{En1=}
\E^{(1)}=
\sum_{i=1}^k
\{(\frac{1}{x_{i+1}}-\frac{1}{x_i})\lambda_i+\frac{2(x_{i+1}-x_i)}{\sqrt{r_i}}
\}\EEA
with $x_{k+1}\equiv 1$ and 
\BEQ 
\Delta_n=(\frac{1}{x_1}-\frac{1}{n})\frac{\lambda_0}{n}+\frac{2(x_1-n)}{n\sqrt{r_0}}
\EEQ
Only the terms in $\Delta_n$  depend explicitly on $n$. Using  
$r_0=r_1+(1/\lambda_0-1/\lambda_1)/x_1$, which follows from (\ref{ri=}),
 the dependence on $\lambda_0$ has 
been made explicit. The saddle point w.r.t. $\lambda_0$ yields
\BEQ
\lambda_0=\frac{n^2x_1^3}{(1+\lambda_0(x_1r_1-\frac{1}{\lambda_1}))^3}
\approx n^2x_1^3(1-3n^2x_1^3(x_1r_1-\frac{1}{\lambda_1}))+\cdots\EEQ
in agreement with previous finding $\lambda_0=-nq_0\sim+n^2$.
We can now expand $\Delta_n$ in powers of $n$
\BEQ\label{Delta}
\Delta_n=x_1^3 -nx_1^2-n^2x_1^6(x_1r_1-\frac{1}{\lambda_1})
+n^3x_1^5(x_1r_1-\frac{1}{\lambda_1})+\cdots
\EEQ
With this form for $\Delta_n$, $\E_n$ has to be optimized in $x_1,\cdots,x_k$
and $\lambda_1\cdots,\lambda_k$. The $n^2$ term of $\E_n$
has prefactor $x_1^3$, which  is expected to vanish for $k\to\infty$.
So infinite replica symmetry breaking cures the problem of negative variances.
We have thus reached the main point of this work. It remains to be shown that 
the leading term (which is of order $n^5$), has a non-vanishing  prefactor.
Before doing that we calculate the average free energy.

\subsection{The average free energy}

Let us now calculate the ${\cal O}(n)$ value of $\tilde E_n$ and later
the leading correction, that will turn out to be of order $n^5$. 

In the continuum limit one gets from eq. (\ref{En1})
\BEQ
\E^{(1)}=\int_0^1dx q(x) +2\int_0^1\frac{dx}{\sqrt{r(x)}}
\EEQ
\BEA r(x)&=&\frac{x}{\lambda(x)}-\int_x^1\frac{dy}{y^2\lambda(y)}
\longrightarrow r'(x)=\frac{q'(x)}{\lambda^2(x)}
\label{r=}\\
\lambda(x)&=&-xq(x)+\int_0^xdyq(y)
\longrightarrow \lambda'(x)=-xq'(x)
\label{lab=}\EEA
We search a solution non-constant $q(x)$ for $0<x<\x$, while $q(x)=\q$
for $\x<x<1$ for some $\x$. Variation with respect to $q(z)$ yields
\BEQ \label{qeq}
\frac{1}{r^{3/2}(z)\lambda^2(z)}-
\int_z^\x\frac{dx}{xr^{3/2}(x)\lambda^2(x)}
-\frac{1}{z\lambda(z)^2}\int _0^z\frac{dx}{r^{3/2}(x)}
+\int_z^\x\frac{dy}{y^2\lambda^2(y)\int_0^y}\frac{dx}{r^{3/2}(x)}=1
\EEQ
Differentiation with respect to $z$ yields $q'(z)=0$ or
\BEQ\label{difff}
-\frac{1}{r^{5/2}(z)\lambda(z)}+
\frac{2z}{r^{3/2}(z)}-2\int_0^z\frac{dx}{r^{3/2}(x)}=0
\EEQ
Taking two more derivatives one obtains
\BEQ
r(z)\lambda(z)=\frac{5}{6z}
\EEQ
Inserting this in eq. (\ref{difff}) one gets an equation for $r$ alone.
This yields
\BEQ 
r(x)=\frac{A}{x^6}\longrightarrow  \lambda(x)=\frac{5x^5}{6A}
\EEQ
with some $A$. Inserting this in eq. (\ref{r=}) one finds
\BEQ \x=\frac{5}{6} \EEQ
and finally from (\ref{qeq}) one obtains
\BEA \label{rx}
q(x)&=&-\frac{2\,3^9x^4}{5^6}\\
r(z)&=&\frac{5^8}{2^43^{10}x^6}\\
\lambda(z)&=& \frac{2^3 3^9 x^5}{5^7}
\EEA
The average free energy is determined by
\BEQ \E^{(1)}=\int_0^{\x}q(x)dx+(1-\x)\q+2\int_0^{\x}\frac{dx}{\sqrt{r(x)}}+
2\frac{1-\x}{\sqrt{\r}}=\frac{27}{100}
\EEQ
This value is approximately $4$ times smaller than the exact value $\pi/3$.

\subsection{Cumulants of the free energy}

Next we consider the higher powers in $n$, coded in eq. (\ref{Delta}) for
$\Delta_n$. For our purpose we may calculate it using the $n=0$ expressions 
for the $x_i$ and $\lambda_i$.
The terms of order $n^0$, $n^1$ and $n^2$ have too many powers of
$x_1$ and vanish for $k\to\infty$, where $x_1\to 0$. The $n^3$ term yields
\BEQ
\E^{(5)}=x_1^6r_1-\frac{x_1^5}{\lambda_1}
\EEQ
Using the continuum results (\ref{rx}) it can be estimated by setting 
$r_1\approx r(x_1)$,
$\lambda_1\approx \lambda(x_1)$
\BEQ \label{Eappr}
\E^{(5)}\approx -\frac{5^7}{2^4 3^{10}}
\EEQ
The correct approach is to calculate $r_1$  and $\lambda_1$ at finite but
large $k$. At $n=0$ this can be done quite simply. Taking derivatives of
(\ref{En1=}) with respect to $\lambda_1$ and $x_1$ we obtain,
\BEQ
\frac{1}{x_2}-\frac{1}{x_1}-\frac{x_2-x_1}{r_1^{3/2}}
\,\,\frac{-1}{x_2\lambda_1^2}=
\frac{\lambda_1}{x_1^2}-\frac{2}{\sqrt{r_1}}=0,
\EEQ
respectively. $x_2$ can be eliminated, and we get 
\BEQ
\lambda_1=8x_1^5 \qquad r_1=\frac{1}{16x_1^6}
\EEQ
We donot need the value of $x_1$ here, and obtain immediately
\BEQ
\E^{(5)}=x_1^6r_1-\frac{x_1^5}{\lambda_1}=-\frac{1}{16}
\EEQ
The prefactor in the estimate (\ref{Eappr}) has the same sign but
is larger by a  factor $1.32305$. This result thus proves the leading
behavior 
\BEQ
\overline {(-\beta F)^5_{cum}}=\frac{5!}{16}L \frac{\gamma\sigma^4}{2\pi}
\EEQ
which has the expected sign. 
In other words, in the variational approach 
free energy fluctuations are predicted to be of order $1/L^{1/5}$.
\section{Summary}
In the present work we have extended the variational approach of
Honeycutt and Thirumalai applied to
directed polymers in 1+1 dimensional random media. This is a toy model for 
the more general field of interfaces in random media, considered by 
M\'ezard and Parisi. In our problem a mapping on a Schr\"odinger equation is 
possible, which prevents the need of space-dependent propagators and
the occurrence of wavenumbers.

We have reconsidered the variational Ansatz of 
Honeycutt and Thirumalai. We have shown
that  it leads to a negative variance of the physical free energy, and thus is
not a correct Ansatz.

Next we have extended this Ansatz by allowing off-diagonal components of the
matrix $q_{\alpha\beta}$. It was observed that an extra constraint is needed
to make the problem well posed. 
The only natural and realistic candidate is spatial 
translational invariance of the variational wavefunction in replica space.
In doing so, we have seen that both the replica symmetric and one-step
replica symmetry broken solutions still lead to a negative variance,
though it has become smaller. We observe that more
and more breakings improves the matter and that for infinite
breaking the variance of the free energy has become zero (to leading order
in the system size $L$). 
Infinite order breaking was also assumed by M\'ezard and 
Parisi. We find that the first non-vanishing cumulant
is the fifth one, 
implying free energy fluctuations of order $L^\chi$ with $\chi=1/5$, which 
agrees with the estimate of M\'ezard and Parisi. Let us recall that this is 
the Flory or mean field estimate; loop effects should modify it to yield
the exact  result $\chi=1/3$. 

This present situation of negative variances is reminiscent of the mean field
model of spin glasses, SK-model, where an infinity of breaking was needed to
obtain a vanishing  zero-point entropy. Both effects are related with
disappearence of unstable modes in the limit of infinite breaking. 
\acknowledgments
D. S. thanks the University of Amsterdam, where part of this work was done,
for hospitality.

\end{document}